\documentclass[aps,preprint,groupedaddress]{revtex4}
\usepackage{amsmath}
\usepackage{amssymb}
\usepackage{graphicx}
\begin{document}

%\preprint{}
%\begin{center}
%\title of paper
\title{THE MINIMUM AND MAXIMUM TEMPERATURE OF BLACK BODY RADIATION}



\author{M. Nowakowski and I. Arraut}

%\thanks{GOD}
\affiliation{Departamento de Fisica, Universidad de los Andes, 
Cra. 1E No.18A-10, Bogota,
Colombia}


%\date{\today}

\begin{abstract}
We show, in different ways, that in the ubiquitous phenomenon
of black body radiation there exists a minimum and maximum
temperature. These limiting values are so small and large respectively,
that they are of no
practical use, except in an extreme situation of black hole evaporation
where they lead to maximum and minimum mass.
\end{abstract}

%\pacs{}
%\keywords{}

\maketitle

In 1966 Andrei Sakharov proved an interesting result. Based on very
general arguments,
he showed that the temperature $T$ in black body radiation is limited
by $T \le {\cal T}_{\rm max} \simeq m_{\rm Pl}=G_N^{-1/2}$ where $G_N$
is Newton's gravitational constant \cite{Sakharov}. 
The value of ${\cal T}_{\rm max}
\sim 10^{32}\,\, K$
is exorbitantly large to be of any use, or so it seemed at least 
for a long time. Ten years after Sakharov's result, Hawking derived 
semi-classically the
temperature-mass ($M$) relation for the 
evaporation of black holes, $T=1/(8\pi G_NM)=m_{\rm Pl}^2/(8\pi M)$,
with the spectrum of black body radiation \cite{Hawking, Page}. 
Meanwhile, the
result of Sakharov was forgotten and so the early
opportunity to make a  connection between the two results got lost. 
However, it is not only obvious that the high temperature 
${\cal T}_{\rm max}$ can be reached only 
in black hole evaporation, it sets also 
a minimum mass $m_{\rm Pl}/(8\pi)$ for the black hole
which is usually called black hole remnant. 
This remarkable fact,
corroborated by semi-classical arguments
in a different context as shown below, can be extended to demonstrate 
the existence of a minimum temperature provided we introduce a cosmological constant
\cite{Carroll} as an explanation of the accelerated universe \cite{acceleration, Knop}.
We first briefly elaborate on the alternate way to establish a black hole remnant. 
To this end, we 
consider a Generalized Uncertainty Principle (GUP) \cite{Adler1, Adler2}.
If $E=p$  is the photon's energy, then the acceleration of a test particle 
at a distance $r$ is
$a_G=\frac{G_NE}{r^2}$.
As an order of magnitude 
estimate, we can write for the displacement due to gravitation
$\Delta x_G\simeq \frac{G_NE}{r^2}L^2\simeq G_NE=G_Np$
where we used  $r \sim L$ ($L$ is a typical length scale entering the
problem).
Setting  $\Delta p \sim p$
one arrives at the GUP relation
\begin{equation}\label{gup3}
\Delta x \ge \frac{1}{2\Delta p} + \frac{G_N \Delta p}{2},
\end{equation}
which generalizes the Heisenberg uncertainty relation by introducing gravity effects within.
This uncertainty relation  has also been derived independently, by different means, in 
\cite{ven1, ven2, Maggiorre, Scar}. 
Identifying $\Delta x$ with the Schwarzschild radius, i.e.,
$\Delta x \sim 2 r_s=2G_NM$
and the energy uncertainty with the temperature,
$\Delta p \sim E \sim T$,
one can establish a relation between $T$ and $M$ via the GUP relation. The result reads
\cite{Adler1, Adler2}
\begin{equation} \label{gup6}
2G_NM=2\frac{M(T)}{m_{\rm Pl}^2}=\frac{1}{2T}+\frac{T}{2m_{\rm Pl}^2}.
\end{equation}
Solving this equation for  $T=T(M)$
and introducing a calibration factor 
$(2\pi)^{-1}$ (we do not expect to get all factors right by invoking
arguments from the quantum mechanical uncertainty relation alone) gives
\begin{equation} \label{gup7}
T=\frac{1}{\pi}\left(M-\sqrt{M^2-m^2_{\rm Pl}/4}\right).
\end{equation}
Two conclusions are in order:
(a) Equation  (\ref{gup7}) reduces to Hawking's radiation formula $T=1/(8\pi G_N M)$ 
for large  $M$. 
(b) To derive it via the GUP relation is a nice and economic way displaying also the main
quantum issues involved.
There is, however, a difference as compared to the standard Hawking formula,
namely, the existence of a black hole  remnant to ensure the existence of a positive  $T$:
$M > {\cal M}_{\rm min}=\frac{m_{\rm Pl.}}{2}$. 
It is worth noting that there exists a connection  between the choice of 
the sign in the solution (\ref{gup7}) and the negative `heat capacity' 
of the
black hole radiation. Indeed, the curve $M(T)$ in (\ref{gup6}) has two 
regions, one with negative slope $dM/dT < 0$, and the other one for higher $T$
with $dM/dT > 0$. The former, corresponding to the choice of
the sign in (\ref{gup7}) and to the existence of
an invertible map $M(T)$, is also the region preferred by 
physical arguments. Hence, 
everything is consistent. The condition
\begin{equation} \label{capacity}
\frac{dM(T)}{dT} < 0
\end{equation}
will be also used later in the text.

Clearly, the GUP result 
can now be interpreted in terms of a maximal temperature confirming thereby Sakharov's result
from
black body radiation.
The values of the minimum mass derived via GUP and black body radiation are
not exactly the same. We would, however, not expect an exact agreement while handling
order of magnitude arguments. However, the fact that both independent ways
lead to the existence of a black hole remnant is remarkable. In times, when there   
is no general consent on quantum gravity, such an agreement from different sources is a valuable
information and a hint that we are on the right track. The consistency of the existence
of a maximum temperature in black body radiation can also be confirmed using 
a second way to derive it \cite{Massa}.
This method is then also suitable to open another doorway discussed below.
Because of the definition of proper time in General Relativity, the $-g_{00}$ component of the metric should be 
positive definite \cite{LL}. We can also regard the mass $M$ entering the Schwarzschild metric
as energy, which in turn, can be replaced by energy density $\rho$, i.e.,
\begin{equation} \label{bb1}
0 < -g_{00}=1-\frac{2G_NM}{R}=1-(8\pi/3)G_N\rho R^2.
\end{equation} 
Hence we have 
$\rho < \frac{3}{8\pi}\frac{1}{G_NR^2}$.
Using the Stefan-Boltzmann law $\rho=\sigma T^4$ gives \cite{Massa}
$T^4 < \frac{3}{8\pi}\frac{1}{\sigma G_N R^2}$.
Finally, to get rid of the radius $R$, we employ the
quantum mechanical result for black body radiation, $R > 1/T$ \cite{Bekenstein, Massa2}. 
The maximal temperature
obtained this way, namely,
\begin{equation} \label{bb4}
T < T_{\rm max}=\sqrt{\frac{45}{2\pi^3}}m_{\rm Pl},
\end{equation}
is of the same order of magnitude as Sakharov's result.
The return of the cosmological constant  $\Lambda$ to explain the 
accelerated stage of the universe makes it a worthwhile undertaking to
look for the effects of $\Lambda$ on the temperature.
Repeating the same steps from above, this time with 
$\Lambda \neq 0$, i.e., using the Schwarzschild-de Sitter metric
we can write
\begin{equation} \label{bb5}
0 < \rho < \frac{3}{8\pi}\frac{m_{\rm Pl}^2}{R^2}-
\frac{1}{8\pi}m^2_{\rm Pl}
m_{ \Lambda}^{2}
\end{equation}
with $m_{\Lambda}=\sqrt{\Lambda}\sim 10^{-29} \,\, {\rm K}$ (we use $\Lambda=8\pi G_N \rho_{\rm vac}$
and the observational value $\rho_{\rm vac} \simeq 0.7 \rho_{\rm crit}$). 
These inequalities can be translated into
\begin{equation} \label{bb6}
\frac{1}{\sqrt{3}}m_{\Lambda}
=T_{\rm min} < T < T_{\rm max}.
\end{equation}
The minimum temperature due to the cosmological constant 
leads via the Hawking formula to a maximum mass of the order 
$m_{\rm Pl}^2/(8\pi m_{\Lambda})$.
Again, we can check this result by a GUP relation including $\Lambda$. 
What we need is the gravitational potential
\begin{equation}
\Phi (r) =-\frac{r_s}{r} -\frac{1}{6}\frac{r^2}{r_{\Lambda}^2}, \,\,\, r_{\Lambda}=
 \frac{1}{\sqrt{\Lambda}}
\end{equation}
and an additional assumption, $\Delta p \sim L^{-1}$, often used in the
context of uncertainty relations \cite{LL2}.
The result is 
\begin{equation} \label{uncertLambda}
\Delta x \ge \frac{1}{2\Delta p} + \frac{\Delta p}{2m_{\rm pl}^2} - \frac{1}{3}
\frac{m_{\Lambda}^2}{\Delta p^3}.
\end{equation}
As in the case of $\Lambda=0$, we can use (\ref{uncertLambda}) 
in analyzing the black hole radiation. 
The steps involved are conceptually equivalent to the ones described above and we quote
only the final result \cite{xxx1}
\begin{equation} \label{gupL1}
M(T)=\frac{m_{\rm Pl}^2}{4}\frac{1}{T}+ \frac{T}{4}- \frac{1}{6}\frac{m_{\rm Pl}^2 m_{\Lambda}^2}{T^3}.
\end{equation}
The curve $M(T)$ 
is schematically depicted in Figure 1 (due to the huge difference between
the values of $m_{\rm Pl}$ and $m_{\Lambda}$ the figure is not drawn to scale, but
it well reflects the main characteristic properties of $M(T)$).
The curve 
$M(T)$ has a zero at ${\cal T}_{0}=\sqrt{2}m_{\Lambda}/\sqrt{3}$.
After passing through zero it rises steeply (positive slope) to a point which is
a local maximum located at 
\begin{equation} \label{ex1}
{\cal T}_{\rm min}=\sqrt{2}m_{\Lambda}.
\end{equation} 
From here on the curve has a negative slope till it reaches a point of a 
local minimum located at 
\begin{equation} \label{ex2}
{\cal T}_{\rm max}=m_{\rm Pl}.
\end{equation}
After the local minimum the curve assumes again a positive slope.
In agreement with equation (\ref{capacity}) we exclude the regions with positive slope and remain
within the domain $T \in [{\cal T}_{\rm min}, {\cal T}_{\rm max}]$.
We can summarize this in one equation, namely,
\begin{equation} \label{gupL3}
 {\cal T}_{\rm max}\sim m_{\rm Pl} \ge T \ge {\cal T}_{\rm min}\sim m_{\Lambda},
\end{equation}
in agreement with the result (\ref{bb6}) found in the context of black body radiation.
The existence of a maximum and minimum temperature automatically guarantees not only a minimum
(remnant) black hole mass, but also a maximum value via,
\begin{equation} \label{mass}
{\cal M}_{\rm min}=M({\cal T}_{\rm max}) \sim m_{\rm Pl} \le M \le 
{\cal M}_{\rm max}=M({\cal T}_{\rm min}) 
\sim M_{\Lambda}=\left(\frac{m_{\rm Pl}}{m_{\Lambda}}\right)m_{\rm Pl}.
\end{equation}

If we accept that the black-hole entropy is given by $S=4\pi \left(\frac{M}{m_{\rm Pl}}\right)^2$, 
we can find the maximum and minimum entropy associated with the maximum and minimum mass, 
respectively. These limiting entropies are given by:
\begin{equation}
S_{BH\;min}\thicksim\pi,\;\;\;S_{BH\;max}\thicksim\left(\frac{m_{\rm Pl}}{m_\Lambda}\right)^2.
\end{equation}
Using the Stephan-Boltzmann law and the result (\ref{gupL3}), 
together with the mass-energy equivalence, 
we can also calculate the maximum and minimum fractional emission rate 
for a black-hole. Defining for 
convenience the quantity $x=\frac{M}{m_{\rm Pl}}$, we obtain for the maximum
value
\begin{equation}   \label{eq:final rate}
\left(\frac{dx}{dt}\right)_{max}\approx-\frac{64}{t_{ch}},
\end{equation}
where $t_{ch}=60(16)^2\pi t_{\rm Pl}$. For the minimum emission rate the expression is,
\begin{equation}
\left(\frac{dx}{dt}\right)_{min}\approx-\frac{2}{\pi}\left(\frac{m_\Lambda^2}{m_{\rm Pl}}\right)
\times 10^{-3}.
\end{equation}
\begin{figure}
\includegraphics[angle=0, width=6cm]{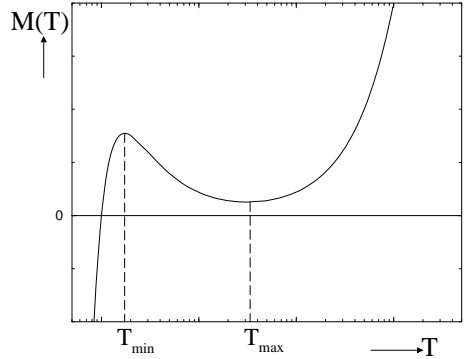}
\caption{Schematic diagram of $M(T)$ defining the minimum and maximum
temperature according to the Generalized Uncertainty Principle.}
 \end{figure}

To summarize, we established the following results regarding:
(i) the existence of a minimal (due to $\Lambda$) and maximal temperature in black body radiation,
(ii) the existence of a black hole remnant of the order of Planck's mass and
(iii) the existence of a maximal black hole mass (due to $\Lambda$). 
We have confirmed these results in different independent ways and 
as far as the order of magnitude is concerned,
all the results are consistent with each other. For instance, 
the results in (\ref{bb4}) 
and (\ref{bb6}) agree
with the estimates obtained in (\ref{ex1}) and (\ref{ex2}).
Whatever the nature of true quantum gravity, these results do not depend
on the details of a quantum gravity theory.
We find this a notable fact. Note also the dual role of the
constants $G_N$ and $\Lambda$ in (\ref{bb6}) and (\ref{gupL3}) encountered also elsewhere
in gravity theory with $\Lambda$ \cite{we1, we2}.

\end{document}